\documentclass[fleqn,usenatbib]{mnras}
\usepackage{amsmath}
\usepackage{graphicx}
\usepackage{amssymb,txfonts}
\usepackage{xspace}

\newcommand{\integral}{{\textit{INTEGRAL}}}
\newcommand{\xte}{{\textit{RXTE}}}

\newcommand{\xmm}{{\textit{XMM-Newton}}}
\newcommand{\sax}{{\textit{Beppo\-SAX}}}
\newcommand{\gro}{{\textit{CGRO}}}
\newcommand{\fermi}{{\textit{Fermi}}}

\newcommand{\swift}{{\textit{Swift}}}
\newcommand{\msun}{{\rm M}_{\sun}}

\newcommand{\g}{$\gamma$}

\newbox\grsign \setbox\grsign=\hbox{$>$} \newdimen\grdimen \grdimen=\ht\grsign
\newbox\simpropbox
\setbox\simpropbox=\hbox{\raise.5ex\hbox{$\propto$}\llap
     {\lower.5ex\hbox{$\sim$}}}\ht2=\grdimen\dp2=0pt

\def\sym{$S_YZ_6R_{30}T_{150}C_2$\xspace}
\def\slm{$S_LZ_6R_{20}T_\infty C_5$\xspace}
\def\ssm{$S_SZ_4R_{20}T_{150}C_5$\xspace}

\title[High-energy gamma-rays from Cyg X-1]{High-energy gamma-rays from Cyg X-1}

\author[A. A. Zdziarski et al.]
{Andrzej A. Zdziarski,$^1$ Denys Malyshev,$^{2}$ Maria Chernyakova$^{3,4}$ and Guy G. Pooley$^5$\\
$^1$Nicolaus Copernicus Astronomical Center, Polish Academy of Sciences, Bartycka 18, PL-00-716 Warszawa, Poland\\
$^2$Institut f{\"u}r Astronomie und Astrophysik T{\"u}bingen, Universit{\"a}t T{\"u}bingen, Sand 1, D-72076 T{\"u}bingen, Germany \\
$^3$School of Physical Sciences, Dublin City University, Glasnevin, Dublin 9, Ireland \\ 
$^4$DIAS, Fitzwiliam Place 31, Dublin 2, Ireland\\
$^5$Cavendish Laboratory, J. J. Thomson Avenue, Cambridge CB3 0HE, UK\\
}

\date{Accepted 2017 July 19. Received 2017 July 19; in original form 2016 July 18}

\pagerange{\pageref{firstpage}--\pageref{lastpage}}
\pubyear{2017}

\begin{document}

\maketitle

\label{firstpage}

\begin{abstract}
We have obtained a firm detection of Cyg X-1 during its hard and intermediate spectral states in the energy range of 40~MeV--60 GeV based on observations by the \fermi\/ Large Area Telescope, confirming the independent results at $\geq$60MeV of a previous work. The detection significance is $\simeq\!8\sigma$ in the 0.1--10~GeV range. In the soft state, we have found only upper limits on the emission at energies $\gtrsim$0.1 MeV. However, we have found emission with a very soft spectrum in the 40--80~MeV range, not detected previously. This is likely to represent the high-energy cutoff of the high-energy power-law tail observed in the soft state. Similarly, we have detected a \g-ray soft excess in the hard state, which appears to be of similar origin. We have also confirmed the presence of an orbital modulation of the detected emission in the hard state, expected if the \g-rays are from Compton upscattering of stellar blackbody photons. However, the observed modulation is significantly weaker than that predicted if the blackbody upscattering were the dominant source of \g-rays. This argues for a significant contribution from \g-rays produced by the synchrotron-self-Compton process. We have found that such strong contribution is possible if the jet is strongly clumped. We reproduce the observed hard-state average broad-band spectrum using a self-consistent jet model, taking into account all the relevant emission processes, e$^\pm$ pair absorption, and clumping. This model also reproduces the amplitude of the observed orbital modulation. 
\end{abstract}

\section{Introduction}
\label{intro}

Cyg X-1, an archetypical black-hole binary, shows two main spectral states, hard and soft. In the hard state, the main component of its X-ray spectrum appears to be thermal Comptonization in a plasma with the electron temperature of $k T_{\rm e}\sim 100$ keV, which features a sharp cutoff [in the $EF(E)$ representation] at energies $E\gtrsim 200$ keV. Beyond $\sim$1 MeV, there is a clear high-energy tail, measured up to $\sim$3 MeV (e.g., \citealt{mcconnell02}, hereafter M02; \citealt*{jrm12,zls12}). The origin of the photon tail may be Compton scattering by a power-law electron tail above the thermal electron distribution in the accretion flow (e.g., M02). In the soft state, there is a strong disc blackbody component in the X-ray spectrum, peaking at $\sim$1 keV, followed by a pronounced high-energy tail, measured up to $\simeq$10 MeV (M02). 

\citet*{mzc13}, hereafter MZC13, detected high-energy \g-ray emission from Cyg X-1 based on observations by the Large Area Telescope (LAT) on board of \fermi. The detection was at a $\simeq\! 4\sigma$ significance, which corresponds to the chance probability of the detection of $\simeq\! 6\times 10^{-5}$ if the noise distribution is Gaussian. Moreover, that emission was present only in the hard spectral state, while only upper limits were found in the soft state. This ruled out the observed source being an artefact of the background subtraction, thus providing a strong argument for the emission actually coming from Cyg X-1.

After the release of the new and much improved LAT calibration, {\sc Pass 8}, in the summer of 2015, and given the significantly increased on-source time with respect to the data analysed in MZC13, we embarked on a new analysis, which results we present here. During our work, the work by \citet{zanin16} (hereafter Z16) appeared. Independently of our results, they have obtained the detection of \g-ray emission from Cyg X-1, only in the hard state, at a $\simeq\! 8\sigma$ level. They also found some evidence for orbital modulation of the flux, expected if the emission originates, at least partly, from Compton scattering of stellar blackbody photons \citep{jackson72}, which process we hereafter abbreviate as BBC. Here, we provide results of our analysis of the emission, which extends the analysis of Z16 by presenting the discovery of strong emission at the softest measured energies in both hard and soft states, quantifying the orbital modulation in the hard state, and developing theoretical models of the \g-ray emission and its modulation. Furthermore, we present 15 GHz monitoring data from the Arcminute Microkelvin Imager (AMI), covering the entire duration of the LAT observations analysed here. The part of these data after MJD 57211 has not been published before.

In our theoretical modelling, we adopt the parameters of Cyg X-1 similar to those in MZC13. The orbital period is $P\simeq 5.6$ d, and we assume the black-hole mass of $M_{\rm X}\simeq 16\msun$, and the (relatively uncertain) mass of the donor as $M_*\simeq 27\msun$ \citep{cn09,orosz11,ziolkowski05,ziolkowski14}, which correspond to the separation between the components of $a\simeq 3.2\times 10^{12}$ cm. We assume the stellar effective temperature of $T_*\simeq 2.55\times 10^4$ K and the luminosity of $L_*\simeq 4.8\times 10^{38}$ erg s$^{-1}$, which are the lower limits in the analysis of \citet{ziolkowski14}. We adopt the jet inclination with respect to the binary plane of $i\simeq 30\degr$ \citep{orosz11, ziolkowski14}, and the distance of $D=1.86$ kpc \citep{reid11}. The projected opening angle of the steady jet, present in the hard state, was constrained by \citet{stirling01} to $\lesssim 2\degr$. Given $i\simeq 30\degr$, the actual opening angle, $\Theta_{\rm j}$, is the above times $\sin i$, leading to $\Theta_{\rm j}\lesssim 1\degr$. We adopt here $\Theta_{\rm j}\simeq 0.5\degr$. The jet velocity is constrained by the lack of the counterjet to $\beta_{\rm j}\gtrsim 0.8$ \citep*{stirling01,zpr16}, and we assume here that lower limit.

\begin{table*}
\begin{center}
\caption{The new LAT sources in the ROI. The tentative identifications are based on the SIMBAD database. The source {\tt n5} does not appear in Z16. }
\begin{tabular}{cccccc}
\hline
Number& TS(3--30~GeV) & RA, Dec & Name in Z16 & Tentative ID & Source type\\
\hline
{\tt n1} & 31 & 302.42, 35.68 & J2009+35 & HD 191612 & spectroscopic binary (O8), \\
&&&&&X-ray source (\xmm, {\it Rosat})\\
{\tt n2} & 18 & 301.24, 34.38 & J2005+34 & PSR J2004+3429 & pulsar\\
{\tt n3} & 27 & 298.69, 33.45 & J1955+33 & PN K 3--49 & planetary nebula\\
{\tt n4} & 24 & 297.26, 34.20 & J1949+34 & V* V1449 Cyg,& dwarf nova, X-ray source\\
&&&&1RXS J194917.1+341042\\ 
{\tt n5} & 31 & 297.60, 34.95  & -- & HD 226099, & spectroscopic binary (G5), \\
&&&&1RXS J194932.7+350119 & X-ray source\\
\hline
\end{tabular}
\end{center}
\label{new}
\end{table*}

\section{Analysis of the \textit{FERMI}\/ data}
\label{analysis}

\begin{figure}
\centerline{\includegraphics[width=\columnwidth]{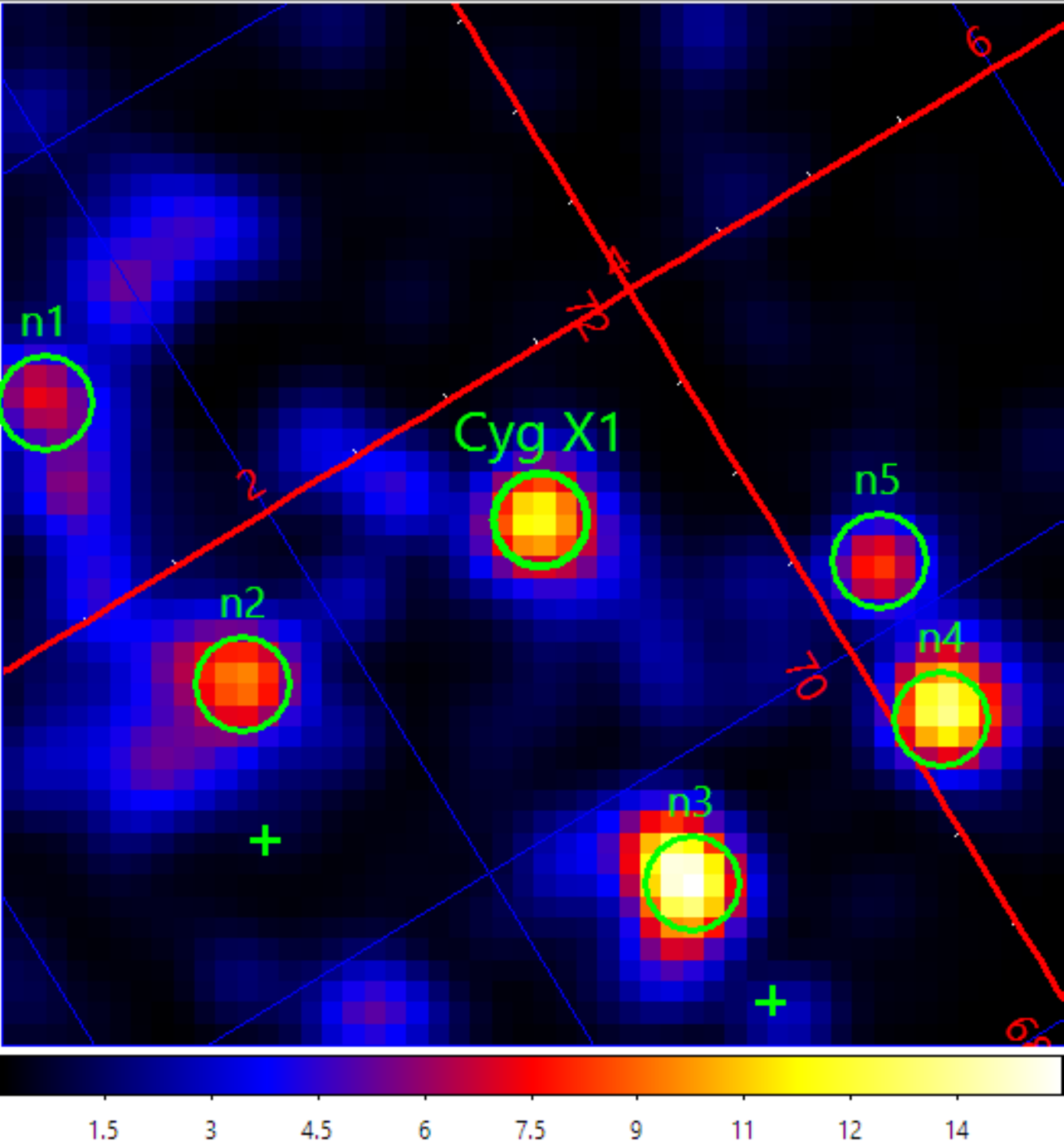}} 
\caption{The TS map (Galactic coordinates) at energies $\geq$3 GeV for the data within a $5\degr\times 5\degr$ square around the position of Cyg X-1 with the 3FGL sources subtracted. We see the presence of a point-like source at the position of Cyg X-1 and 5 new point-like residuals, marked as {\tt n1}--{\tt n5}.
} \label{map}
\end{figure}

\begin{table}
\begin{center}
\caption{The adopted intervals of the hard and intermediate states analysed by us in MJD during the \fermi/LAT observations, following the \fermi\/ launch on MJD 54628. MJD 57705 is the last day of the analysed data. }
\begin{tabular}{cc}
\hline
Start & End\\
\hline
54628 &55390\\
55665 &55795\\
55895 &55940\\
56035 &56090\\
56735 &56750\\
56760 &56845\\
57012 &57045\\
57105 &57265\\
57330 &57705\\
\hline
\end{tabular}
\end{center}
\label{dates}
\end{table}

We have analysed the available \fermi/LAT data (MJD 54682--57705) coming from the direction of Cyg X-1 using the latest version of the Fermi Science Tools (v10r0p5) with the P8R2\_CLEAN\_V6 instrument response function (IRF) and the standard value of the zenith angle cut of $z_{\rm max}=90\degr$. In order to reduce a possible contribution from the Earth limb emission at $E\lesssim 200$~MeV, we have applied a stricter choice of $z_{\rm max}=80\degr$. However, we have found this reduction with respect to $90\degr$ has a relatively minor effect on our results. In order to account for the broad \fermi/LAT point-spread function (PSF) at all energies studied by us (40~MeV--300~GeV), we consider a broad, $25\degr \times 25\degr$ region of interest (ROI) around the Cyg X-1 position. We have included in the modelling all sources within the ROI from the 4-year (3FGL) Fermi catalogue \citep{3rdcat} as well as the standard templates for the Galactic (\texttt{gll\_iem\_v06.fits}) and extragalactic (\texttt{iso\_P8R2\_CLEAN\_V6\_v06.txt}) diffuse backgrounds. The catalogue sources were assumed to be described by power law spectra with the fitted indices frozen to the best-fit values over the whole considered energy range, while the normalizations were considered to be free. In Appendix \ref{soft}, we compare the results using the above method with that in which the ROI is increased by $10\degr$ and the normalizations of the sources there are kept at the 3FGL values, and find this change affects our results relatively little. We note that the background templates do not cover the lowest energies, $\leq$60~MeV, which we also analyse. Our method to deal with that issue is described in detail in Appendix \ref{soft}. The spectral analysis has been performed with the python tools\footnote{\url{fermi.gsfc.nasa.gov/ssc/data/analysis/scitools/python\_tutorial.html}.} provided by the \fermi/LAT collaboration. The upper limits are calculated with the \texttt{UpperLimits} python module for TS (test statistic; see \citealt{mattox96}) $<4$ detection significances, which correspond to a 95 per cent ($\simeq 2\sigma$) probability for the flux to be lower than that specified.

At the initial stage of our analysis, we built the TS map of the Cyg X-1 vicinity in the 3--300~GeV energy band, see Fig.\ \ref{map}. This map clearly reveals the presence of a point-like source at the position of Cyg X-1 (with TS $\simeq 16$, corresponding to a $\sim\! 4\sigma$ detection significance) and 5 new point-like residuals, marked on the map as {\tt n1}--{\tt n5}. We summarize the available information on those sources in Table \ref{new}, including the positions (as the positions of local TS maxima at the map) and the TS(3--30~GeV) values. We find the {\tt n1} source to be of a particular interest as its direction coincides with an O8 high-mass binary, detected by both \xmm\/ and {\it Rosat}, implying it may be a new \g-ray loud binary (see, e.g., \citealt{dubus13} for a review of this class of sources). Note that all 3FGL sources were subtracted from this map. Hereafter, we include these new sources in the spectral model of the ROI.

We then divide the available LAT observations into the hard/intermediate and soft state. The inclusion of the intermediate state in the former category is motivated by the finding that the radio flux from Cyg X-1, which is seen to originate from the jet \citep{stirling01}, is correlated with the X-ray flux in both the hard and intermediate states and it achieves its overall maximum in the latter. The decline of the radio emission takes place only in the soft state \citep{z11b}. The high-energy \g-ray emission studied by us most likely also originates from the jet.

We perform the division based on the light curves from the \xte\/ All-Sky Monitor (ASM; \citealt*{brs93,levine96}), MAXI \citep{matsuoka09}, and AMI. The AMI Large Array is the re-built and reconfigured Ryle Telescope. \citet{pf97} describe the normal operating mode for the Ryle telescope in the monitoring observations; the observing scheme for the AMI Large Array is very similar. The new correlator has the centre frequency of 15 GHz and a useful bandwidth of about 4 GHz (compared to 15 GHz and 0.35 GHz, respectively, for the Ryle). 

From fig.\ 3 in \citet{z11b}, we find that the radio flux in the soft state becomes weaker than that in the hard state and starts to decline fast when the photon spectral index, $\Gamma(3$--12\,keV) becomes $\gtrsim$2.5. Here, $\Gamma(3$--12\,keV) is defined based on the ASM fluxes in the 3--5 keV and 5--12 keV channels, using the method described in \citet*{zps11}. We have found that for contemporaneous ASM and MAXI data, the index $\Gamma(2$--10\,keV) based on the 2--4 keV and 4--10 keV channels of the MAXI data is lower by $\simeq$0.1 on average than $\Gamma(3$--12\,keV). Thus, we use here the criterion of $\Gamma(3$--12\,keV$)\lesssim 2.5$, $\Gamma(2$--10\,keV$)\lesssim 2.4$ to identify our hard/intermediate state. This yields the hard state intervals as given in Table \ref{dates}. The remaining intervals are defined by us to represent the soft state. The LAT exposures for our data are 143.4 Ms and 118.4 Ms in the hard and soft state, respectively. 

\begin{figure*}
\centerline{\includegraphics[width=17.1cm]{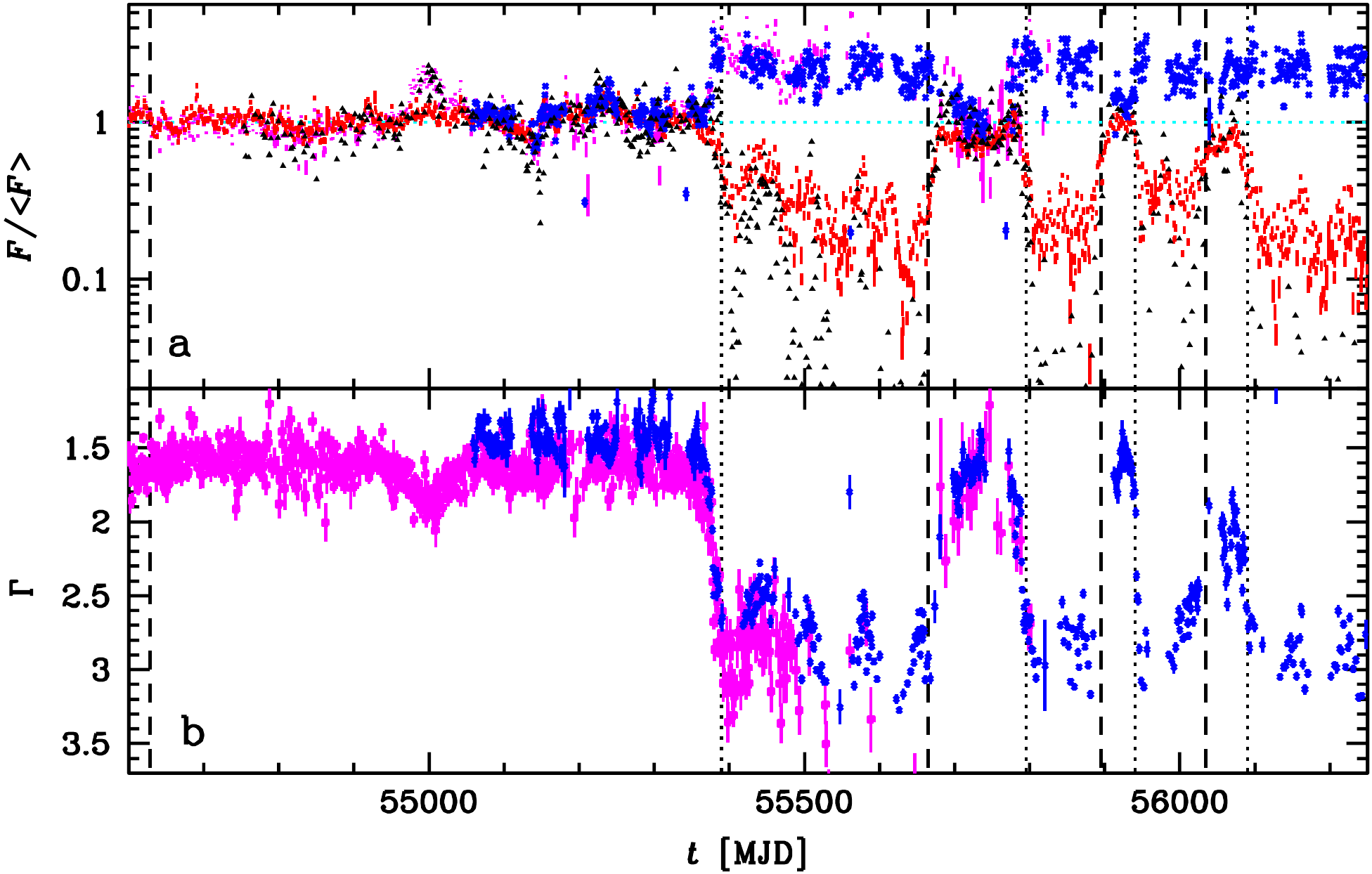}} 
\centerline{\includegraphics[width=17.1cm]{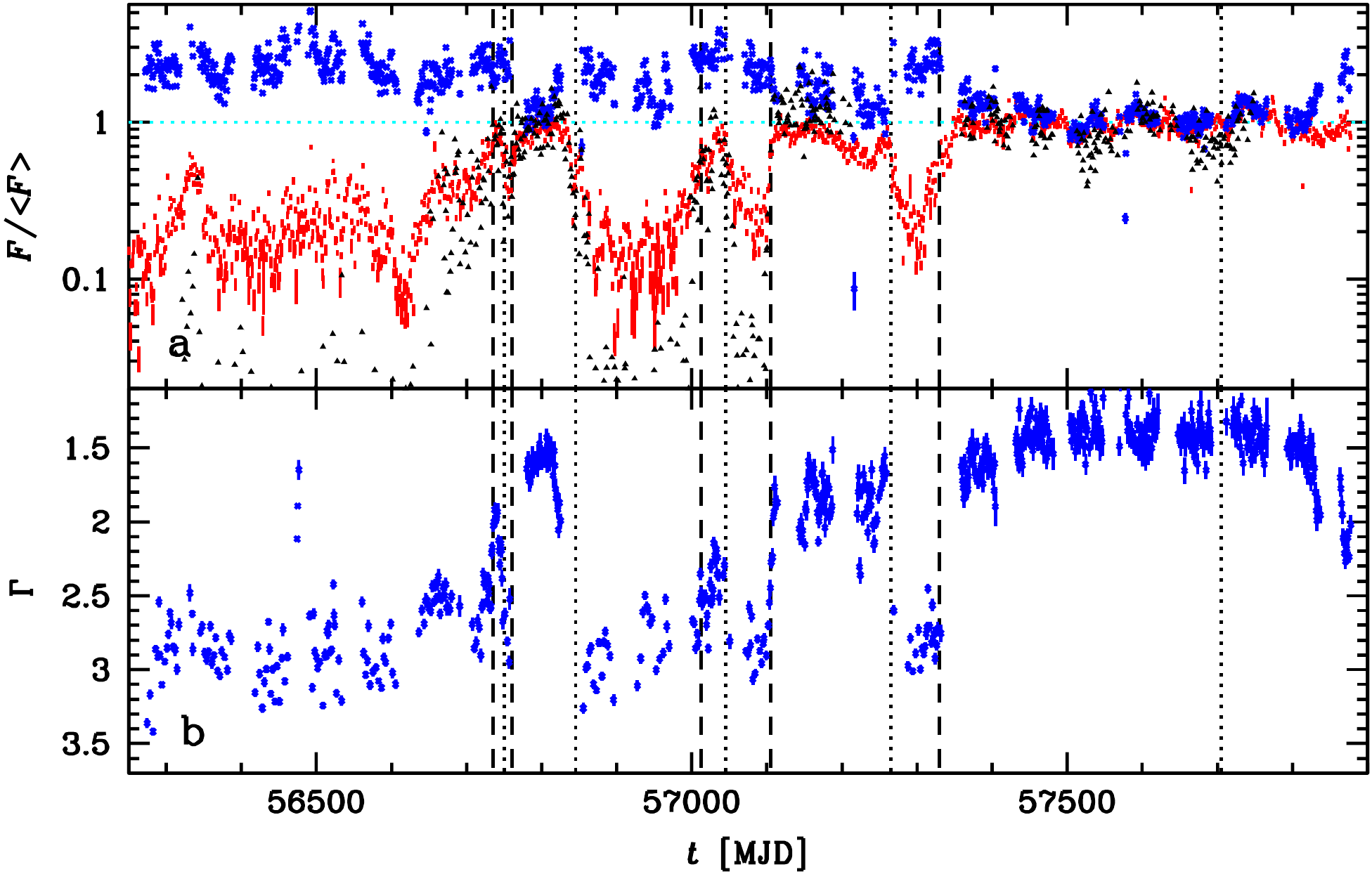}} 
\caption{(a) Light curves of Cyg X-1 from the ASM (1.3--12 keV, squares with error bars, magenta), BAT (15--50 keV, circles with error bars, red), MAXI (2--20 keV, crosses with error bars, blue), and AMI (15 GHz, triangles w/o error bars, black) normalized to their respective average hard-state values (dotted horizontal line, cyan) of $\langle F\rangle \simeq 20.7$ s$^{-1}$, $\simeq 0.173$ cm$^{-2}$ s$^{-1}$, $\simeq 1.0$ cm$^{-2}$ s$^{-1}$, and $\simeq 11.6$ mJy, respectively. The vertical dashed and dotted lines denotes the start and end, respectively, of the analysed hard/intermediate state intervals (Table \ref{dates}). The first dashed line and the last dotted line show the range of the studied LAT data. (b) The spectral indices in the 3--12 keV (squares with error bars, magenta) and 2--10 keV (crosses with error bars, blue) ranges based on the ASM and MAXI data, respectively.
} \label{lc}
\end{figure*}

The time dependencies of the fluxes and indices for the period studied here are shown in Fig.\ \ref{lc}. We also show the 15--50 keV fluxes from the \swift\/ Burst Alert Telescope (BAT; \citealt{barthelmy05,m05}). We normalize the ASM, BAT and AMI count rates/fluxes to the average values during the long hard state of MJD 53880--55375 \citep{z11b}. In the case of MAXI, we normalize its light curve to the hard-state average during the overlap with the ASM. 

\begin{figure}
\centerline{\includegraphics[width=\columnwidth]{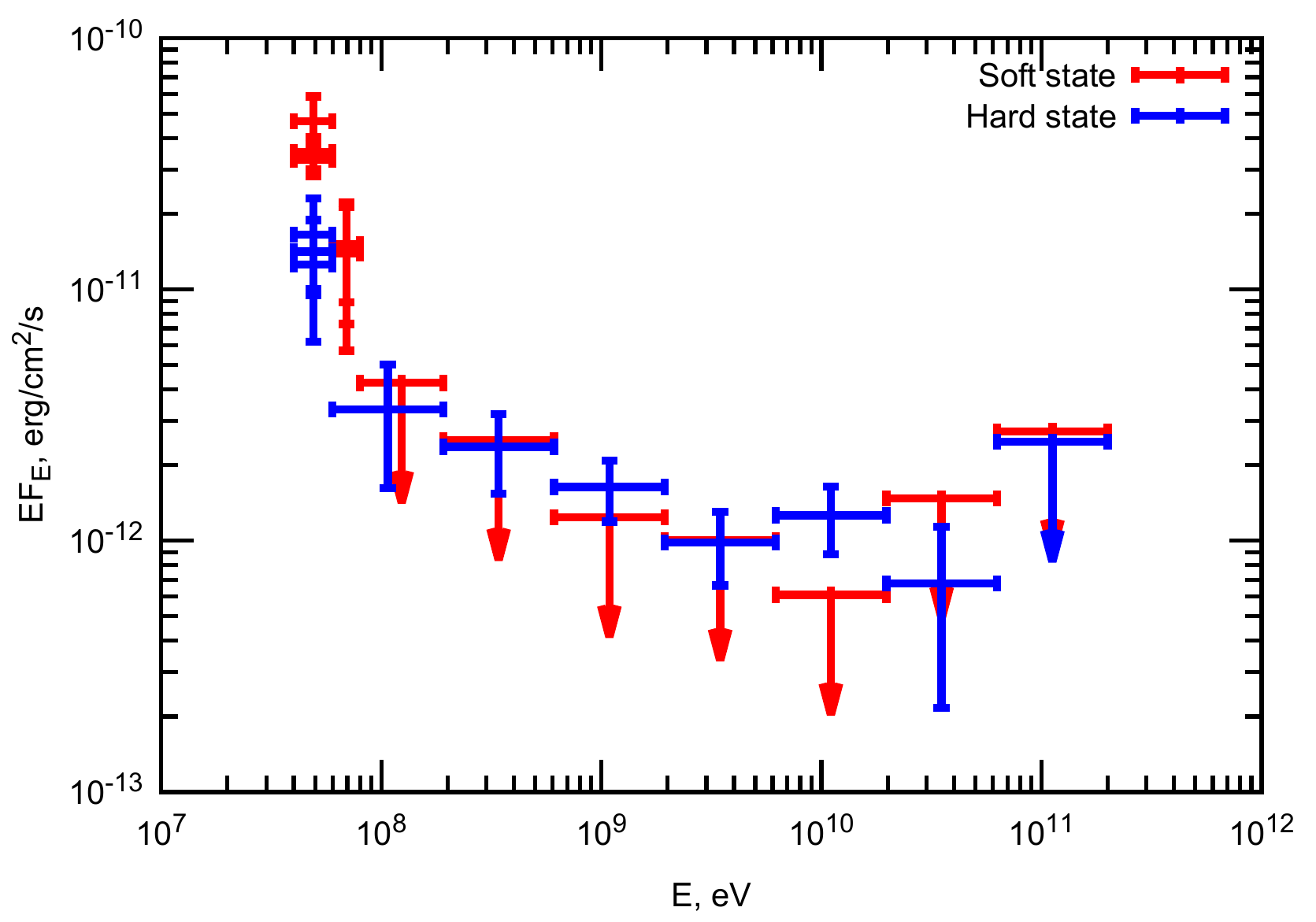}} 
\caption{The \fermi\/ LAT measurements and upper limits in the hard/intermediate state (blue error bars with crosses) and the soft state (red error bars). The triple points in the 0.04--0.0.08 MeV range correspond to the three considered background models, namely \sym, \ssm and \slm, in the order of the increasing flux. This illustrates the expected level of systematic uncertainties at these energies, see Appendix \ref{soft}.
} \label{spectra}
\end{figure}

We have clearly detected Cyg X-1 in the hard/intermediate state in the 0.04--60 GeV range, which we show in Fig.\ \ref{spectra}. The 0.1--10 GeV detection significance is of $>8\sigma$, confirming MZC13 and Z16, and the spectrum in this range can be fitted as a power law with the photon index of $\Gamma\simeq 2.4\pm 0.2$. The 0.1--60 GeV flux is $1.1\pm 0.2\times 10^{-11}$ erg cm$^{-2}$ s$^{-1}$, corresponding to the luminosity of $4.6\pm 1.0\times 10^{33}$ erg s$^{-1}$ at $D=1.86$ kpc and assuming isotropy. However, we have also detected Cyg X-1 in the soft state with a very soft spectrum at lowest measured energies, 40--80~MeV, which is a new result. At higher energies in the soft state, we have obtained upper limits, more restrictive than those in MZC13, see Fig.\ \ref{spectra}. 

\begin{figure*}
\centerline{\includegraphics[width=14cm]{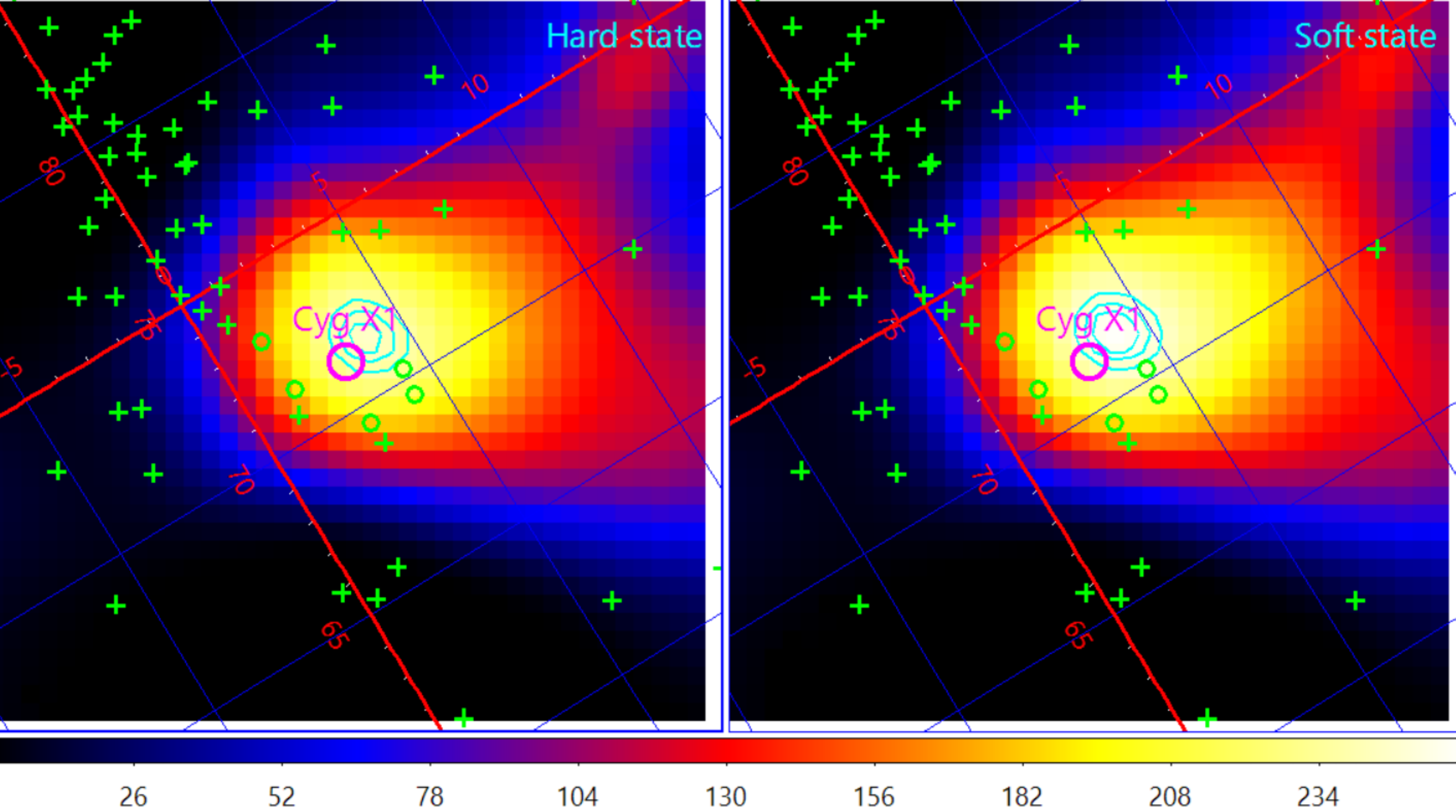}} 
\caption{The TS maps at energies of 40--60 MeV for the data within a $20\degr\times 20\degr$ square (pixel size $0.5\degr\times 0.5\degr$) around the position of Cyg X-1 in the Galactic coordinates for the hard (left panel) and soft (right panel) states. The 3FGL and {\tt n1}--{\tt n5} sources have been subtracted, and their positions are shown by green crosses and circles, respectively. The countours show $1\sigma$, $2\sigma$ and $3\sigma$ uncertainties around the positions of the TS maxima.
} \label{map4060}
\end{figure*}

\begin{figure}
\centerline{\includegraphics[width=\columnwidth]{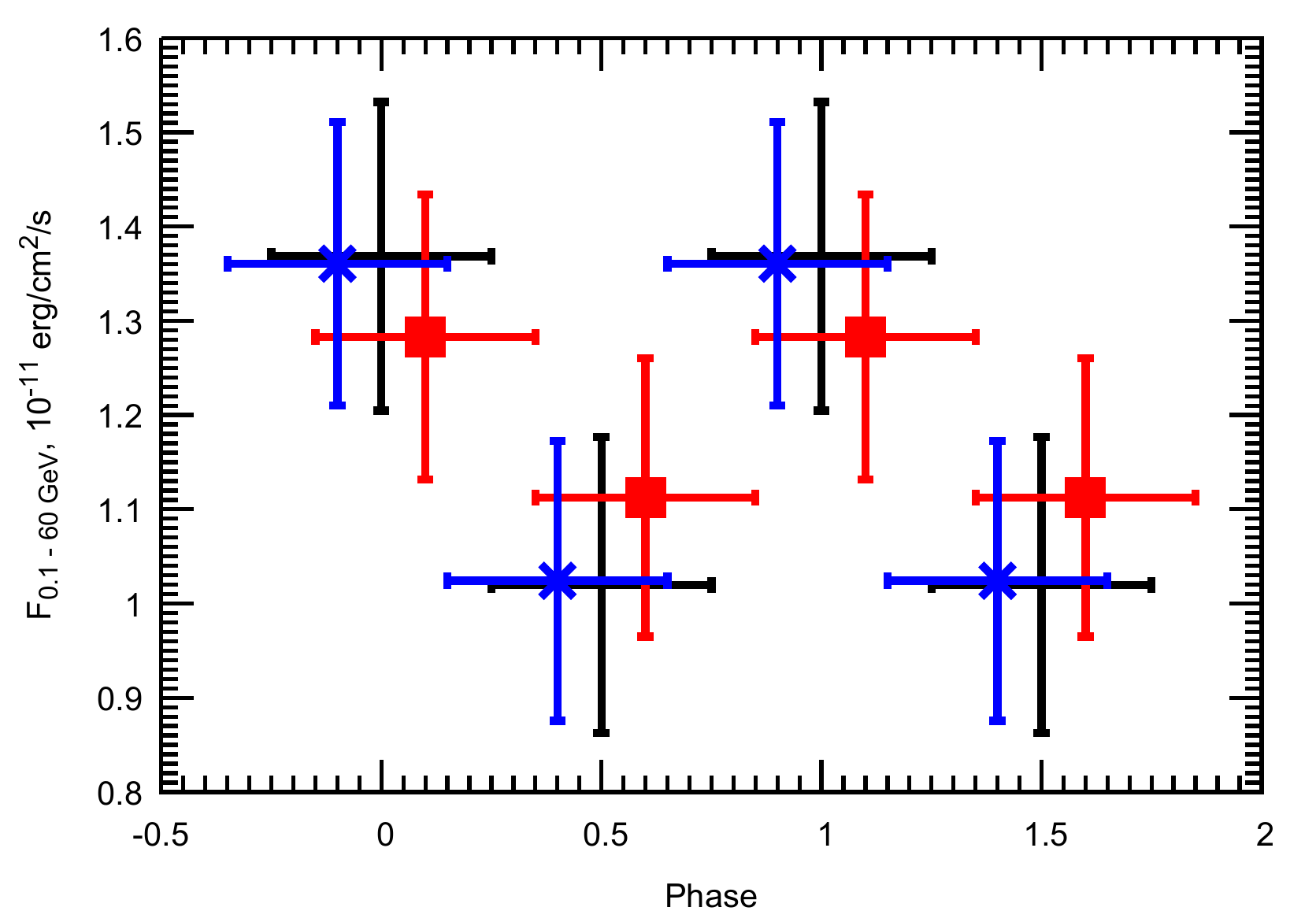}} 
\caption{The orbital modulation in the hard state in the 0.1--10 GeV range, with the orbital phase divided into two bins. For clarity, two full phase ranges are shown. We show the results for the bins centred on the phases of 0, 0.5 (black error bars), $(-0.2, 0.3$; crosses with blue error bars) and (0.2, 0.7; filled squares with red error bars), respectively. We see that the negative offset has a very minor effect, while the positive one causes the modulation to almost disappear.
} \label{orbital}
\end{figure}

At energies $\geq$60~MeV, our results are almost identical to those of Z16, except that we have detected Cyg X-1 in the soft state in the 60--80 MeV range.  In the 40--80 MeV energy range, we show three flux ranges for each state, which correspond to the three considered background models (see Appendix \ref{soft}). This shows the expected level of systematic uncertainties at these energies. We note the high formal detection significance of Cyg X-1 at those energies in both states, with TS(40--60 MeV) $\simeq 180$--280, TS(60--80 MeV) $\simeq 23$--28 in the soft state, and TS(40--60MeV) $\simeq 120$--220 in the hard state. The given TS ranges cover the results from the three assumed Galactic background models, see Appendix \ref{soft}. However, the \fermi/LAT IRF is poorly known at these energies, where the PSF can be as broad as $\gtrsim\! 10\degr$. Together with the overall crowdedness of Cyg X-1 region, this may lead to the source confusion problem, in which case the lowest energy \fermi/LAT spectral points rather show the overall level of the emission in Cyg X-1 region and may, in principle, be also interpreted as upper limits. 

On the other hand, the soft-state fluxes at $E\lesssim 80$ MeV are significantly higher than those in the hard state, in the pattern opposite to that at higher energies. This variability correlated with the spectral states of Cyg X-1 would not be present at all if it originated from background/source confusion effects, and it strongly argues for the origin of the measured fluxes from Cyg X-1. In order to quantify the statistical significance of this flux difference, we have calculated the log-likelihood of the hard-state data model with the flux fixed it its best-fit soft-state value, $L_1$, and that for the flux as a free parameter, $L_2$. The difference, $L_1-L_2$, is distributed as $\chi^2$ with 1 d.o.f.~\citep{wilks38}. We have found the values of $L_1-L_2$ at the 40--60~MeV energy range to correspond to $\simeq 2.8$--$4\sigma$ significances (for different choices of the Galactic diffuse background template, see Appendix \ref{soft}). 

Another piece of evidence for the origin of the observed low-energy emission from Cyg X-1 is provided by the 40--60~MeV TS map of the region ($20\degr\times 20\degr$), shown in Fig.~\ref{map4060}. For both the soft and hard state, the location of TS maximum is consistent with the position of Cyg X-1. The observed $\sim$(1--2)$\sigma$ discrepancy can originate either from statistical fluctuations or from poor modelling of the astrophysical background at low energies.

We have then looked into a dependence of the hard-state emission on the orbital phase. Such a dependence is expected if a substantial part of the emission is due to the BBC process \citep*{jackson72,dch10,z14a}. We have used the ephemeris of \citet{brocksopp99}. Given the limited photon statistics available, we have divided the orbital phase, $\phi$, into two parts only. We have found the 0.1--10 GeV energy fluxes in the orbital phases of $-0.25<\phi/2\upi<0.25$ (around the superior conjunction) and $0.25<\phi/2\upi<0.75$ (around the inferior conjunction) of $1.36\pm 0.17\times 10^{-11}$ and $1.02\pm 0.16\times 10^{-11}$ erg cm$^{-2}$ s$^{-1}$, respectively. Thus, a modulation appears to be present, but still at a relatively low statistical significance. We have also looked at a possible offset of the maximum of the flux with respect to the zero orbital phase. Such an offset would appear if the \g-rays are emitted by the jet inclined with respect to the normal of the orbital plane. Such a misalignment will appear if the jet is aligned with the black-hole rotation axis, which may be not be exactly perpendicular to the orbital plane. A similar offset, with the maximum of the folded light curve at a $\phi/2\upi<0$, was observed in Cyg X-3 \citep{fermi,dch10}. We thus have moved the middle point of each orbital bin by $\Delta(\phi/2\upi)=\pm 0.1,\,\pm 0.2$. We have found that moving the bins to negative values has almost no effect on the observed modulation, for both $-0.1$ and $-0.2$, but a shift toward positive values causes the modulation to become much weaker, especially for the offset of $+0.2$. We show our results for $\Delta(\phi/2\upi)=0$, $\pm 0.2$ in Fig.\ \ref{orbital}. This result suggests an offset of the peak modulation towards a negative value and asymmetric shape, similar to the case of Cyg X-3. Given the limited statistics available, we are unable to determine this shape in detail. We have also searched for a possible dependence of the strength of the orbital modulation on energy, comparing, in particular, the modulation in the 0.1--1 GeV and $>$1 GeV ranges. However, we have found no statistically significant differences.

An orbital modulation during the hard state was also found by Z16. They presented their results as maps of the Cyg X-1 region in the hard state at the phase intervals of $-0.25<\phi/2\upi<0.25$ and $0.25<\phi/2\upi<0.75$. The former map showed Cyg X-1 much more clearly. They give the flux normalization in the bin around the superior conjunction about 50 per cent larger than that for the other bin, which is similar to our result. 

\begin{figure*}
\centerline{\includegraphics[width=15cm]{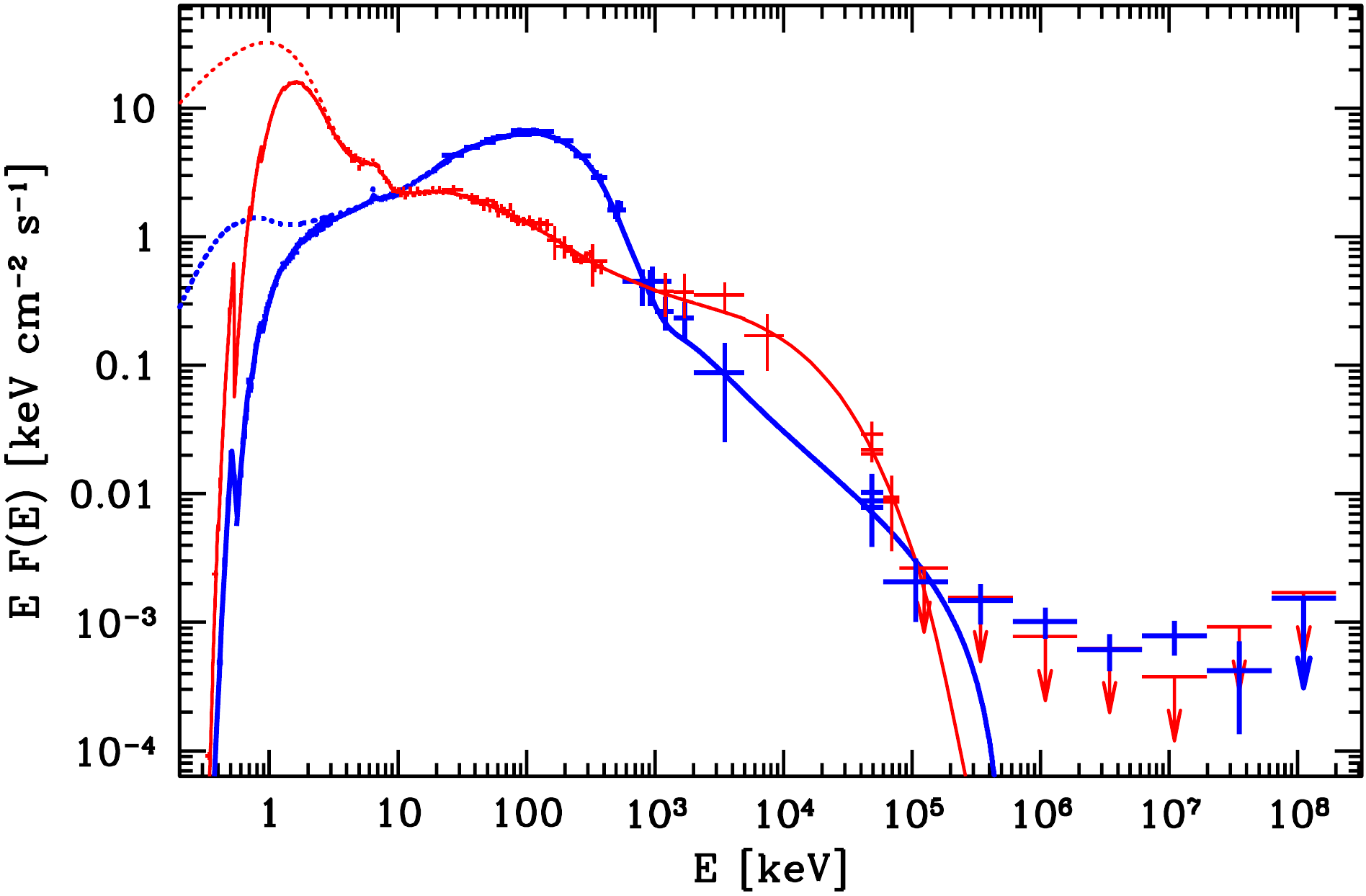}} 
\caption{Broad-band X-ray/\g-ray data for Cyg X-1 in the hard (heavy symbols, blue) and soft (thin symbols, red) states compared to hybrid-Comptonization accretion-flow models. The data at $<$10 MeV (attenuated by X-ray absorption) are from \sax\/ \citep{disalvo01,frontera01} and \gro\/ (M02), and the data at $\geq 40$ MeV are from \fermi/LAT (the same as in Fig.\ \ref{spectra}). The $<$10 MeV data were fitted by hybrid Comptonization using the models of M02 in the hard state and of \citet{pv09} in the soft state. Note that these models have not been fitted to the \g-ray data and yet they predict well the current measurements in that energy range. The heavy (blue) and thin (red) dotted curves at soft X-rays show the unabsorbed models in the hard and soft state, respectively. 
} \label{accretion_model}
\end{figure*}

\section{Theoretical models}
\label{models}

\subsection{Accretion}
\label{accretion}

MZC13 presented a study of the implications of the LAT upper limits and detections for models of accretion flows. Here, we reconsider their analysis using our current measurements. We consider here only leptonic models, and refer to MZC13 for a critical discussion of the hadronic ones. 

Fig.\ \ref{accretion_model} shows the broad-band spectra in X-rays to \g-rays in the hard (blue symbols) and soft (red symbols) states. The data at $\geq$40 MeV are from \fermi/LAT, and are the same as those shown in Fig.\ \ref{spectra}. The data at $\leq$10 MeV are from {\it Compton Gamma-Ray Observatory\/} (\gro; M02) and \sax\/ \citep{disalvo01,frontera01}. Those data show high-energy tails extending to $E\gtrsim 1$ MeV in both hard and soft states of Cyg X-1. Such tails have been detected by \gro\/ (M02) and \integral\/ (e.g., \citealt{jrm12,zls12}). We note here that the hard X-ray/soft \g-ray spectrum in the hard state of \citet{rodriguez15}, who used the \integral\/ data from the so-called Compton mode, is much harder than those of M02, \citet{zls12} and \citet{jrm12}. On the other hand, \citet{laurent16} have found that Compton-mode spectra published earlier need to be revised, resulting in a reduction of the flux values by up to an order of magnitude. Also, the recent \integral\/ results of \citet{walter16} show the hard-state spectrum at $E\gtrsim 1$ MeV significantly weaker and softer than that of \citet{rodriguez15}. Therefore, we do not show it here.

\begin{figure*}
\centerline{\includegraphics[width=16.5cm]{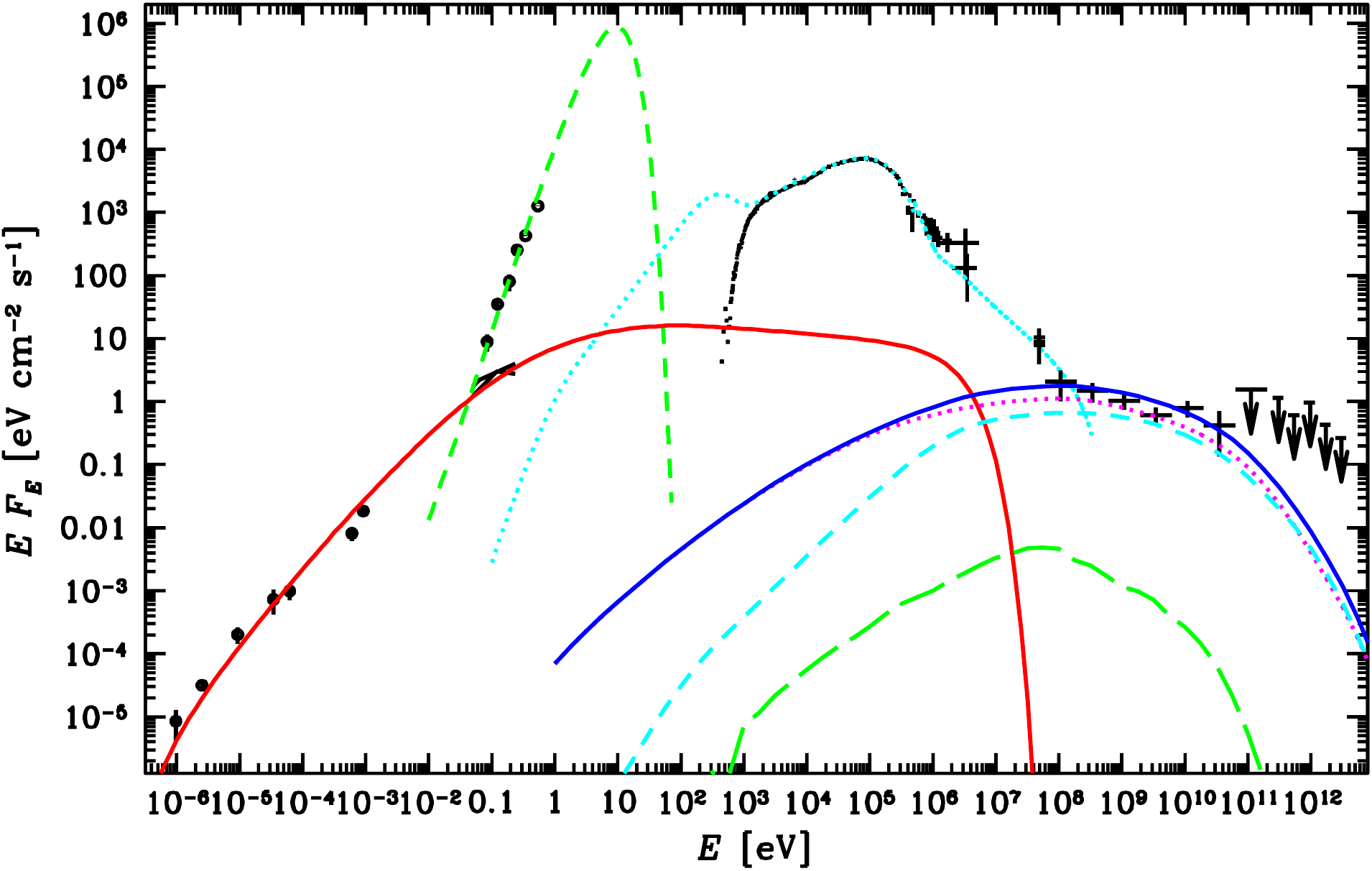}} 
\caption{The average hard-state radio to \g-ray spectrum (points and error bars, black) of Cyg X-1 and its donor shown together with model spectra. The data up to 5 MeV are the same as those in Z14, the 40 MeV--200 GeV symbols give the results of this work, see Fig.\ \ref{spectra}, and the 5 upper limits at the highest energies are from the MAGIC Cherenkov Telescope \citep{magic}. The short-dashed (green) curve shows the stellar blackbody. The dotted (blue) curve shows an unabsorbed accretion disc and hot flow model, see Z14, except that at $E\geq 50$ keV we show the hard-state model of M02. The jet model has the electron injection index of $p=2.2$, see Section \ref{jet}. The solid (left, red), dotted (magenta), long-dashed (green) and dashed (cyan) curves show the model components due to synchrotron, synchrotron self-Compton, and upscattering of disc emission and blackbody-Compton, respectively. The solid curve on the right-hand side (blue) gives the sum of the Compton spectra. 
} \label{jet_model}
\end{figure*}

As discussed in MZC13, the high-energy tails are well modelled by hybrid Comptonization in the accretion flow (e.g., \citealt{av85,pc98,coppi99,gierlinski99}; M02; \citealt{pv09}). A description of that model relevant to our work is given in MZC13. In the hard state, the hybrid plasma probably forms an inner part of the accretion flow, overlapping with the optically thick disc \citep*{dgk07}. In the soft state, the hybrid plasma probably forms coronal regions above an inner part of an optically-thick accretion disc (e.g., \citealt{gierlinski99,dgk07}). In Fig.\ \ref{accretion_model}, we show the fits using the hybrid-Compton model by M02 in the hard state and by \citet{pv09} in the soft state. We see that although both fits were obtained without any knowledge of the spectra at $E>10$ MeV, they do reproduce the 40--200 MeV data well. We consider this coincidence as a further argument for the reality of the detection of Cyg X-1 at $E\lesssim 0.1$ GeV. In the hard state, there is clearly another, harder, component at higher energies, which we discuss in Section \ref{jet} below. 

In the models, high-energy electrons are injected with a power law distribution, and the index of $p_{\rm inj}=2.0$ and 2.2 in the hard and soft state, respectively. In both cases, the maximum Lorentz factor of the injected electrons is $\gamma_{\rm max}=10^3$. The photon spectra are modified by absorption in photon-photon collisions of \g-rays mainly with blackbody photons emitted by the disc, and producing e$^\pm$ pairs. The degree of absorption is controlled by the compactness parameter, $\propto L/R$, where $L$ is the source luminosity and $R$ is the characteristic size. In the hard and soft states, the size of the plasma assumed by M02 and \citet{pv09} is $R\simeq 1.3$, $4\times 10^7$ cm, which corresponds to $\sim$6 and $20 R_{\rm g}$, respectively, where $R_{\rm g}=GM_{\rm X}/c^2$ is the gravitational radius. The observed cutoffs are caused by the pair absorption.

\subsection{Jet}
\label{jet}

Here, we consider jet models for the hard state only. Such models were studied by MZC13 in the case of a fixed electron distribution, following \citet{bk79} and \citet{zls12}. Given the strong radiative cooling present, in particular due to Compton scattering on stellar blackbody photons, this is an unrealistic assumption. Self-consistent models with electron cooling for Cyg X-1 were considered by \citet{z14b}, hereafter Z14, using the formalism developed in \citet{z14a}. Those models assume the profiles of the electron acceleration and magnetic field strength based on physical arguments and calculate the steady-state electron distribution through the jet and the jet emission from the synchrotron, synchrotron self-Compton (SSC), BBC and Compton scattering of photons produced by the accretion flow. Finally, the spectra are attenuated by e$^\pm$ pair absorption, calculated along the jet taking into account the finite size of the donor star.

Fig.\ \ref{jet_model} shows the observed average radio-to-\g-rays spectrum of Cyg X-1 in the hard state. We also show the accretion model of M02, the same as in Fig.\ \ref{accretion_model}. As shown on both Figs.\ \ref{accretion_model} and \ref{jet_model}, it predicts well the lowest-energy points from the LAT. Our jet model should then account for the radio-IR spectrum and the \g-ray spectrum at $E\gtrsim (0.1$--0.2) GeV.

We find here one major change is required with respect to the models of Z14. The \g-ray emission of all those models was dominated by Compton scattering of stellar blackbody photons (BBC), while other contributions, in particular the SSC, made only minor contributions. Thus, these models predict strong orbital modulation of the dominant blackbody-scattering component, by a factor of several, see, e.g., fig.\ 11 of Z14. This is incompatible with the observed relatively weak orbital modulation, see Fig.\ \ref{orbital}. 

We note here that a similarly weak (at face value) orbital modulation is seen during active periods of Cyg X-3, see fig.\ 3 of \citet{fermi}, in which the peak of the \g-ray folded light curve was at a level of only $\sim$1.3 of the minimum. However, most of the emission at the minimum level was attributed by \citet{fermi} to a separate constant flux component. This allowed \citet{dch10} to model that emission entirely in terms of Compton scattering of blackbody photons. In principle, a similar situation may take place in Cyg X-1. We see in Fig.\ \ref{spectra} that the upper limits in the soft state are at relatively similar levels to the hard-state detections. Thus, there could be a persistent component of a different nature at a level just below the minimum of the orbital modulation, which would imply the actual orbital modulation in the hard state to be much stronger.

However, given that this issue cannot be resolved with the available data, we have searched for jet models giving low orbital modulation. Apart from the BBC component, there are also components due to scattering of accretion photons and SSC. The former depends on the accretion flux, which is relatively well constrained by X-ray observations, so there is not much freedom to increase that component. On the other hand, the ratio between the BBC and SSC fluxes depends on the ratio of the densities of the blackbody and synchrotron photons. (In order to minimize this ratio, we have assumed the minimum stellar luminosity compatible with observations, see Section \ref{intro}.) We have searched over a large grid of models yielding the radio and \g-ray fluxes as observed in Cyg X-1, but have found that all of them have the BBC component dominating over SSC. This effect is due to the very large luminosity of the OB supergiant donor in the system.

However, an effect that increases the relative importance of SSC vs.\ external Compton is clumping \citep{stawarz04}. We consider physical conditions in which the accelerated electrons within the jet form $N$ clouds filling a fraction $f$ of the volume. Simple calculations show that this changes the SSC flux (while leaving the synchrotron and external Compton fluxes unchanged) by a factor of $f_{\rm cl}=f^{-2/3}N^{-1/3}$. Thus, we need a small number of high-density clumps filling a small fraction of the jet volume for this effect to substantially increase the SSC flux. We have found that we need $f_{\rm cl}\sim 10^2$, i.e., a strongly clumped jet, for the SSC flux to be comparable to the blackbody Compton one. 

Given the relatively weak observational constraints, we have found a number of models yielding similar radio and \g-ray spectra, see also discussion in Z14. We show our best hard-state model in Fig.\ \ref{jet_model}. The model parameters are as follows. The relativistic electrons are accelerated/injected above the Lorentz factor of $\gamma_{\rm min}=300$ with the index of $p_{\rm inj}=2.2$. (Note that this power-law is then steepened by radiative losses.) The acceleration begins at the distance of about $100 R_{\rm g}$ from the black hole, and the magnetic field strength at that point is $8\times 10^4$ G. See Z14 and \citet{z14a} for a full description of the model. The SSC component at high-energy \g-rays is slightly stronger than the BBC one, and scattering of disc photons is negligible. 

\begin{figure}
\centerline{\includegraphics[width=\columnwidth]{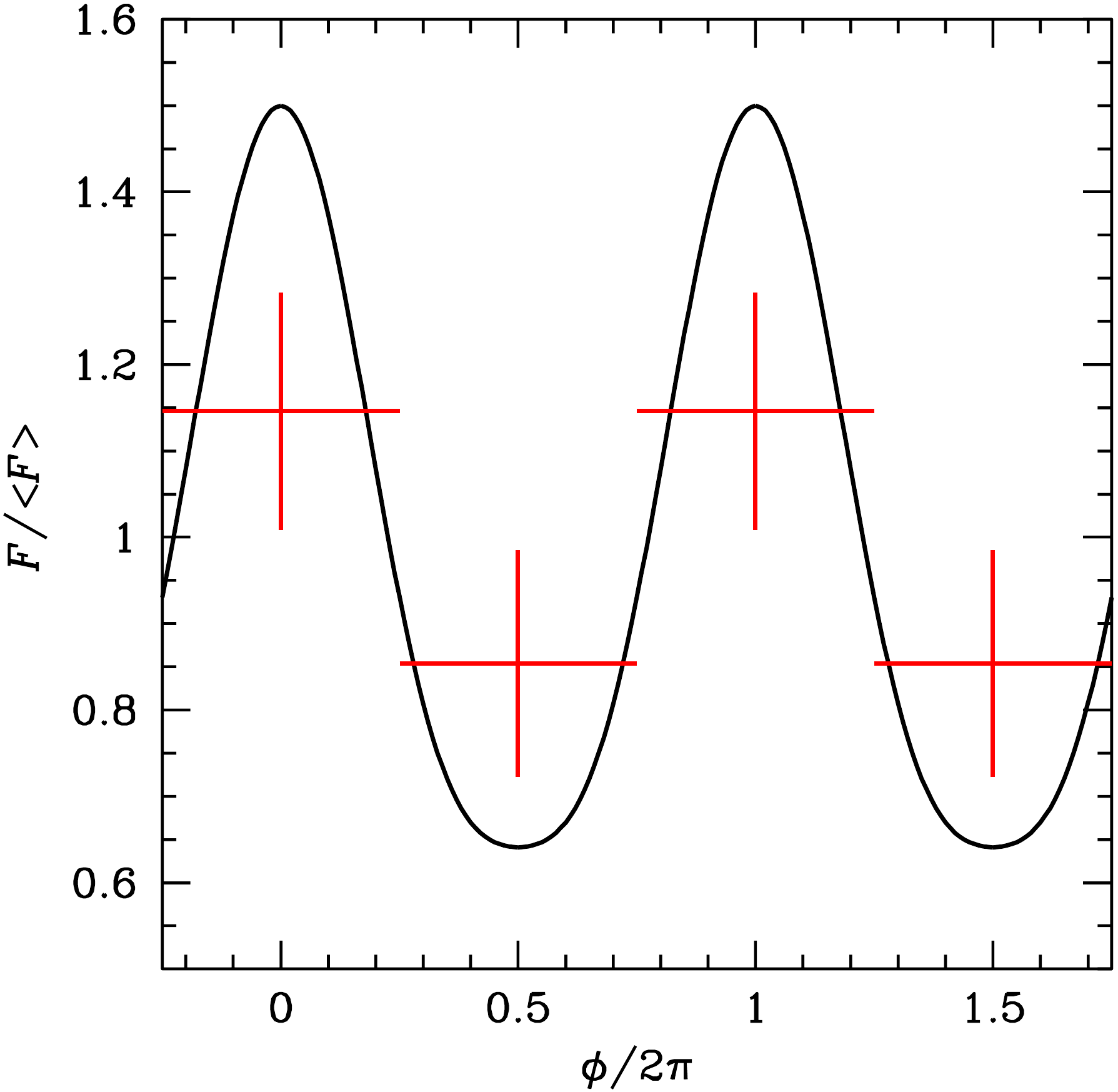}} 
\caption{The observed 0.1--10 GeV orbital modulation in the hard state (error bars) compared to the dependence predicted by the model shown in Fig.\ \ref{jet_model}.} \label{orbital_model}
\end{figure}

The orbital modulation predicted by the jet model is shown in Fig.\ \ref{orbital_model}. It is slightly stronger than that of the best fit, but compatible the observed one within errors. We note that the adopted model treats the stellar blackbody emission as a point source (as in \citealt{dch10}), while the stellar radius is as large as $\sim$1/2 of the binary separation. Taking this effect into account would significantly reduce the predicted modulation \citep*{dch08}. In addition, the jet may precess, though the presence and magnitude of this effect has not yet been constrained \citep{stirling01, rushton09, rushton11}. Precession will change the phases of the maximum and minimum fluxes, which will further reduce the average orbital modulation. Also, variability of the emission will reduce the averaged orbital modulation, due to the dependence of the profile of the \g-ray emission on the mass flow rate through the jet \citep{z14a}. 

\section{Conclusions}
\label{conclusions}

We have obtained measurements and upper limits of the flux from Cyg X-1 in the 0.04--200 GeV energy band based on observations of the \fermi/LAT. Our observational results confirm, and significantly extend, the independent results of \citet{zanin16}. We have detected a steady emission in the 0.04--60 GeV energy band in the hard and intermediate spectral state. The 0.1--10 GeV emission, with the $8\sigma$ significance, can be approximately described as a power law with $\Gamma\simeq 2.4\pm 0.2$. In the soft spectral state, we have found upper limits at $E\gtrsim 80$ MeV, but we have detected a steep soft spectrum in the 40--80 MeV range. The measured 40--80 MeV flux is larger in the soft state than in the hard state (in the pattern opposite to that at higher energies), which argues against its origin from the local background. We discuss the issues related to the analysis at the lowest energies in Appendix \ref{soft}.

We have found that the detections at $E\lesssim$0.1 GeV are well explained by the high-energy tails of the emission of the accretion flow, in both the hard and the soft state. The used models were published in 2002 and 2009 based on data at $E<10$ MeV only, but they still predict well the respective present measurements. This agreement further supports the reality of the LAT detection at the lowest energies. The measured spectra are relatively steep, and correspond to the high-energy cutoffs of the tails caused by the e$^\pm$ absorption.

We have also quantified the orbital modulation of the \g-ray flux. We have not found any statistically-significant dependence of the modulation strength on energy at $E\geq 0.1$ GeV. The peak of the modulation was found at the orbital phases between about $-0.2$ and 0. The observed modulation is significantly weaker than that predicted if the blackbody upscattering were the dominant source of \g-rays. This argues for a significant contribution from \g-rays produced by the synchrotron-self-Compton process. We have found that such strong contribution is possible if the jet is strongly clumped. 

We have reproduced the observed hard-state average radio and \g-ray spectrum at $E\gtrsim 0.1$ GeV using a self-consistent jet model, taking into account all the relevant emission processes, e$^\pm$ pair absorption, and clumping. This model also reproduces the amplitude of the observed orbital modulation.

\section*{ACKNOWLEDGMENTS}

We thank {\L}ukasz Stawarz for discussions and the referee for valuable suggestions. This research has been supported in part by the Polish National Science Centre grants 2012/04/M/ST9/00780, 2013/10/M/ST9/00729, 2015/18/A/ST9/00746, and by the Carl-Zeiss Stiftung through the grant ``Hochsensitive Nachweistechnik zur Erforschung des unsichtbaren Universums" to the Kepler Zentrum f{\"u}r Astro- und Teilchenphysik at the University of T{\"u}bingen. The authors thank the SFI/HEA Irish Centre for High-End Computing (ICHEC) for the provision of computational facilities and support.

\appendix

\section{The analysis of the lowest-energy data}
\label{soft}

A major part of our analysis relies on the standard \fermi/LAT templates for the Galactic diffuse and the extragalactic isotropic backgrounds. Here, we shortly summarize the details on these backgrounds and the non-standard templates used for the low-energy ($\lesssim$80~MeV) part of the analysis.

\begin{figure*}
\centerline{\includegraphics[width=\textwidth]{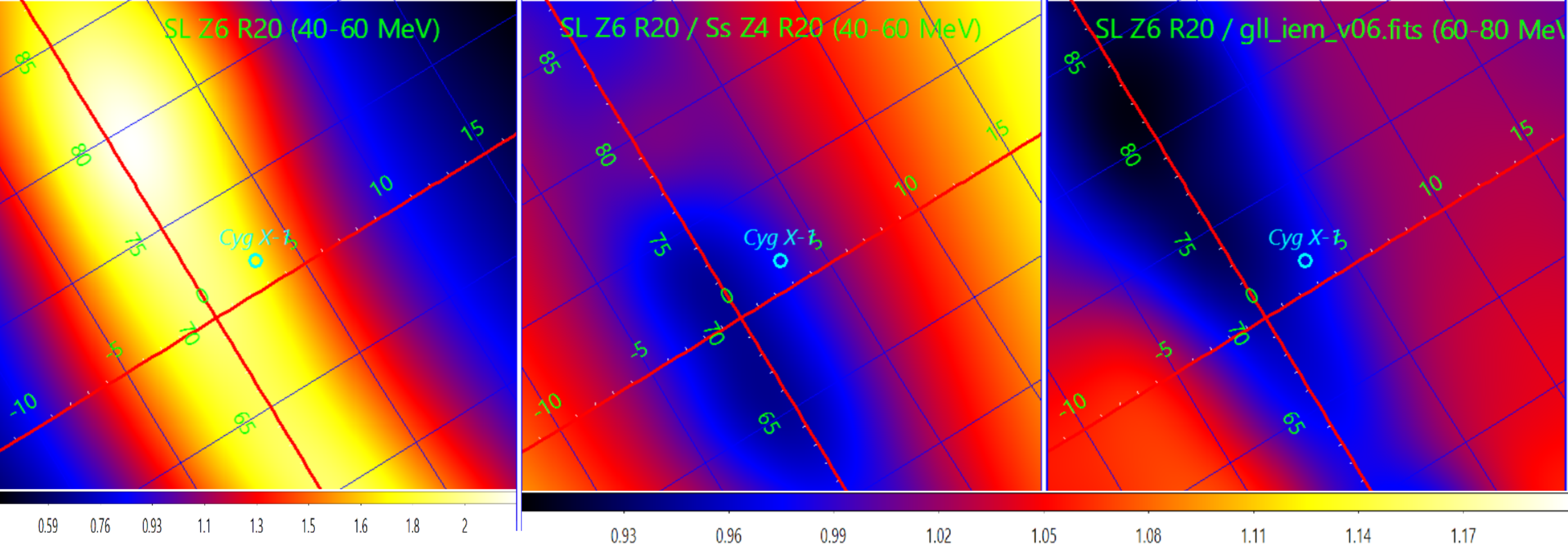}} 
\caption{Left panel: the theoretical count-rate map for the \slm model in the 40--60~MeV band (in arbitrary units). Middle panel: the ratio of the count-rate map for \slm to that for the \ssm model in the 40--60~MeV band. Right panel: the ratio of the \slm count-rate map to the count map expected from the standard \fermi/LAT Galactic diffuse template at the energy range of 60--80~MeV. Note that the standard \fermi/LAT Galactic diffuse background (\texttt{gll\_iem\_v06.fits}) is defined only for energies $\gtrsim$60 MeV. The similarity of variations level allows us to infer that the overall shape of the Galactic diffuse background at 30--80~MeV does not change by more than $\sim$15 per cent and the considered templates could serve for the estimation of the systematics connected with poor knowledge of the exact Galactic diffuse background shape. See text for the details.
} \label{fig:models}
\end{figure*}

The standard \fermi/LAT template for the Galactic diffuse background is phenomenologically based \citep{acero16} on a linear combination of: \\
-- inverse Compton (IC) intensity template predicted by the GALPROP code \citep{GALPROP} for its $S_YZ_6R_{30}T_{150}C_2$ model;\\
-- templates for the H column density (defined in 9 Galactocentric annuli), \\
-- templates accounting for the emission from large-scale diffuse structures (Loop I, Fermi Bubbles, North Polar Spur).  \\
These components in a linear combination with:\\
-- an isotropic template, accounting for the unresolved extragalactic $\gamma$-ray sources and for residual cosmic-ray contamination in the photon data; \\
-- templates of the Solar, Lunar and Earth limb emissions;\\
-- templates for point-like and extended sources from the 3FGL catalogue; \\
were fitted to the all-sky survey data, which allowed the LAT team to determine the coefficients of the above described combination.

In the latest release of the Fermi Science Tools (v10r0p5), the standard template for the Galactic diffuse background (\texttt{gll\_iem\_v06.fits}) is defined at energies of 58.5~MeV--0.5~TeV, while the isotropic template for CLEAN event class (\texttt{iso\_P8R2\_CLEAN\_V6\_v06.txt}) is given for the broader energy range of 34~MeV--0.8~TeV. The analysis below or above these energy thresholds requires the invoking of alternative background models. Following the recommendation of the \fermi/LAT collaboration, we perform the analysis with the enabled energy dispersion handling\footnote{\url{http://fermi.gsfc.nasa.gov/ssc/data/analysis/documentation/Pass8_edisp_usage.html} }.

The non-standard, low-energy, part of our analysis, is performed in the 40--60~MeV and 60--80~MeV energy bands. This is required also for the latter band because the energy dispersion correction, necessary for analysis of the low-energy data, results in the actual bin sizes broader than the nominal ones, which then requires the background templates to be defined in the broader bins. The extent of the broader bins is defined by the energy resolution of \fermi/LAT, known to be 20--30 per cent at energies $\lesssim 100$~MeV\footnote{\url{http://www.slac.stanford.edu/exp/glast/groups/canda/lat_Performance.htm}.}. Thus, for the 60--80~MeV bin, despite the standard Galactic diffuse background available at $\geq$58.5~MeV, the energy dispersion correction algorithms require the knowledge of the background from about 40--50~MeV, which does not allow the use of the standard template. Similarly, for the 40--60~MeV band, the standard isotropic (extragalactic) template cannot be used because we used the broader 30--80 MeV bin, accounting for the energy dispersion.

\begin{figure}
\centerline{\includegraphics[width=\columnwidth]{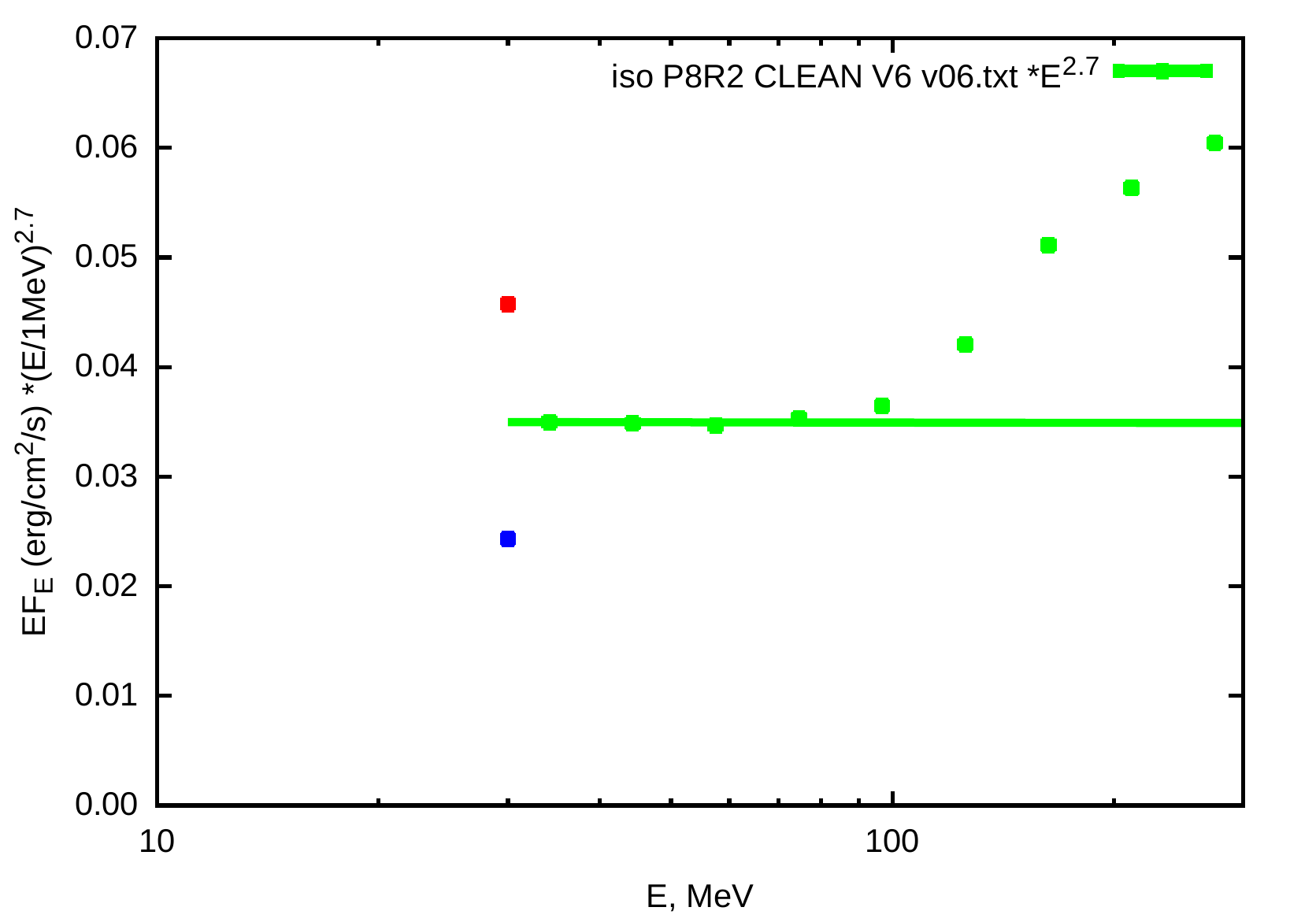}} 
\caption{The spectrum of the isotropic \fermi/LAT background (points at $E\geq 34$ MeV, green) compared to the best-fit power-law model at $E<100$~MeV (with $\Gamma_{\rm iso}=2.7$). The flux of the isotropic background is multiplied by $E^{2.7}$, so the best-fit model corresponds to the horizontal line. The assumed variations of the slope of the isotropic background below 34 MeV are shown with the upper (blue) and lower (red) points at 30 MeV, which correspond to multiplying the flux from the extrapolation of the measured background at $\geq$34 MeV by 0.7 and 1.3, respectively. 
} \label{fig:iso}
\end{figure}

\begin{table*}
\begin{center}
\caption{ The best-fit fluxes (in the unit of $10^{-11}$ erg cm$^{-2}$ s$^{-1}$) for the 40--60~MeV energy bin for the \ssm, \slm and \sym Galactic-background models and different 30--34 MeV fluxes of the extragalactic background. The models iso, iso$-$ and iso$+$ correspond to the extrapolation of the $\Gamma_{\rm iso}=2.7$ power-law extrapolation of the \fermi/LAT isotropic background and that multiplied by 0.7 and 1.3, respectively, see Fig.~\ref{fig:iso}. }
\begin{tabular}{lcccccc}
\hline
Model &   \ssm         & \slm          & \sym        & \ssm           & \slm  & \sym \\
      &   Soft         & Soft          & Soft        & Hard           & Hard  & Hard \\
\hline
iso   & 3.5 $\pm$ 0.5  & 4.6 $\pm$ 1.2 & 3.3$\pm$0.5 & 1.4 $\pm$ 0.5  & 1.7 $\pm$ 0.7 & 1.3$\pm$0.6  \\
iso$-$& 3.6 $\pm$ 0.4  & 4.6 $\pm$ 1.0 & 3.5$\pm$0.5 & 1.8 $\pm$ 0.5  & 1.7 $\pm$ 0.6 & 1.5$\pm$0.6 \\
iso+  & 3.3 $\pm$ 0.5  & 4.2 $\pm$ 1.3 & 3.1$\pm$0.7 & 1.1 $\pm$ 0.7  & 1.2 $\pm$ 0.6 & 1.0$\pm$0.7 \\
\hline
\end{tabular}
\end{center}
\label{fluxes}
\end{table*}

The absence of the standard background templates for low energies has motivated us to use the following modified and extrapolated templates. For the Galactic background, we have used three different templates, \ssm, \slm, \sym, based on the GALPROP code \citep{GALPROP}. These templates were also used by the \fermi/LAT collaboration for the development and testing of the standard template\footnote{\url{http://fermi.gsfc.nasa.gov/ssc/data/access/lat/Model_details/FSSC_model_diffuse_reprocessed_v12.pdf} and \citet{fermi_back}. }. Using three different templates allows us to verify the robustness of the results at the low energies. 

Note that, in addition to the IC template used for the production of standard Galactic background, the GALPROP code also provides physically motivated templates for pion decay and bremsstrahlung emissions based on the H distribution maps and models of cosmic ray production/diffusion in the Galaxy. We have thus added together the templates for the IC, pion-decay, and bremsstrahlung emissions, which results in a single template map, still relatively similar to the standard \fermi/LAT diffuse background (which neglects the pion and bremsstrahlung contributions).

Template maps based on the \ssm and \slm models were also used by the \fermi/LAT collaboration for the comparison and improvement of the standard diffuse background (which uses the template of \sym; \citealt{fermi_back}), and at tested energies are known to describe the data reasonably well. The count-rate map expected from the \slm template (in arbitrary units) at 40--60~MeV in the $25^\circ \times 25^\circ$ around the Cyg X-1 position is shown in Fig~\ref{fig:models}, left panel. The ratio of this map to the map expected from the \ssm template and the one from the standard diffuse background are shown in the middle and right panels of Fig.~\ref{fig:models}, respectively. Fig.~\ref{fig:models} shows that \slm and \ssm template maps vary by $\sim$15 per cent at 40--60 MeV, similarly to the variations at 60--80~MeV with respect to the standard diffuse background. This allows us to assume that the uncertainty of the Galactic diffuse background intensity at 30--80~MeV does not exceed $\sim$20 per cent and the considered \ssm, \slm, \sym templates could serve for the estimation of the systematics connected with poor knowledge of the exact shape of this background.

The isotropic, extragalactic, diffuse background template (\texttt{iso\_P8R2\_CLEAN\_V6\_v06.txt}) is defined as a spatially uniform map with the spectrum which is well described below 100~MeV by a power-law model with the photon index $\Gamma_{\rm iso}= 2.7$, as shown by the horizontal green line in Fig.~\ref{fig:iso}. In our low-energy analysis, we used a power-law extrapolation of the spectrum of this background from 34 MeV down to 30~MeV. To be conservative, we have also considered cases with the 30--34 MeV flux of the isotropic background of 0.7 and 1.3 of the above extrapolation, as shown in Fig.~\ref{fig:iso}. 

The results of the fitting for the 40--60~MeV energy bin for the \ssm, \slm and \sym models and different 30--34 MeV fluxes of the isotropic background are given in Table A1. Although systematically correlated with the assumed 30--34 MeV flux of the isotropic background spectrum, the obtained results show clear detections for all considered models of the isotropic/Galactic diffuse backgrounds. 

\begin{figure}
\includegraphics[width=0.98\columnwidth]{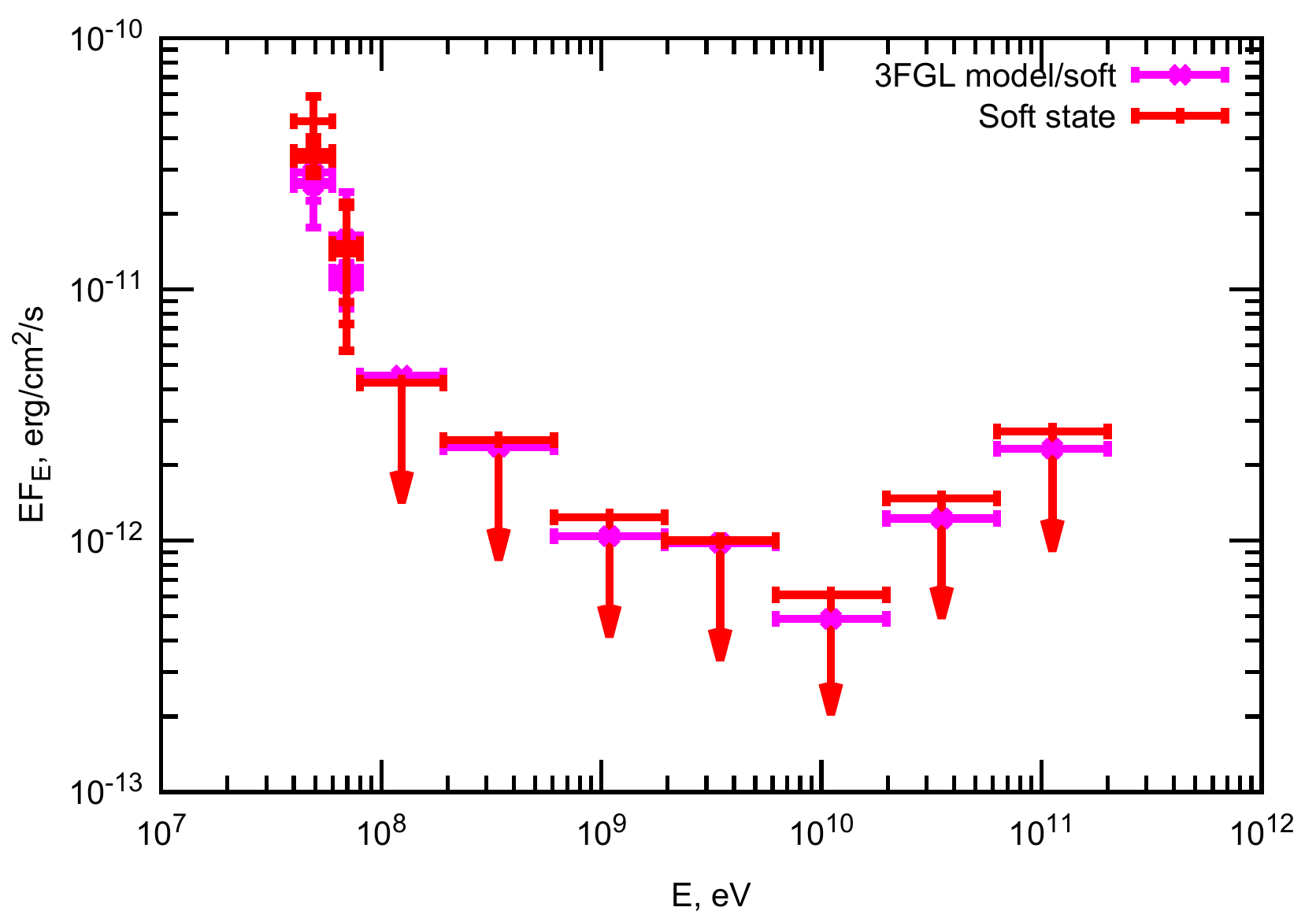}
\includegraphics[width=0.98\columnwidth]{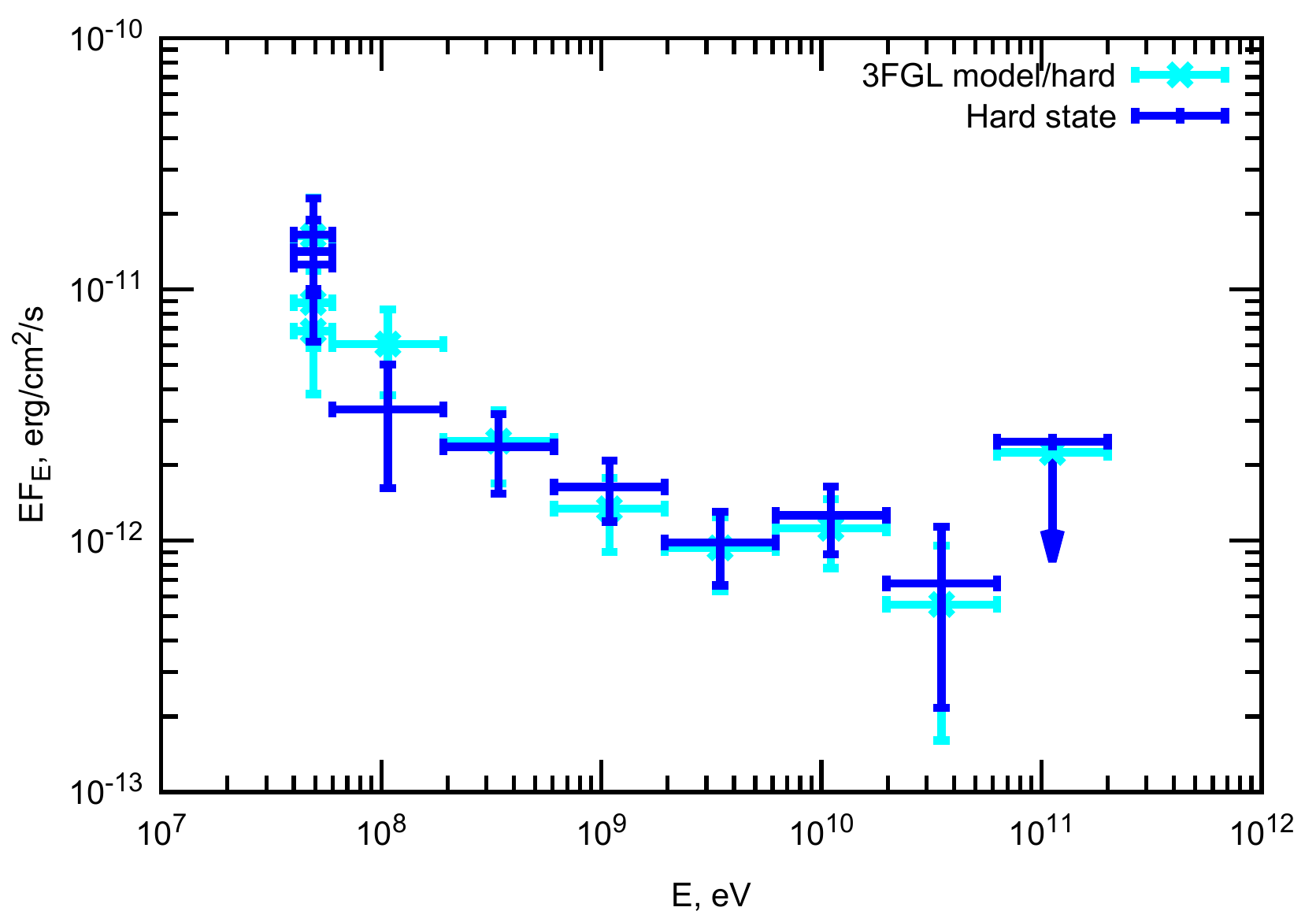}
\caption{The spectra of Cyg X-1 in the states for different choices of the modelling. The red and blue points show the results presented in the main part of the paper, while the magenta and cyan points marked with crosses show the spectra for the ROI model in which the spectral shapes of the sources were fixed to their 3FGL catalogue values and sources from a $10\degr$ region beyond the ROI were added with all the parameters fixed. At energies $\leq$80 MeV, we show the results obtained with the three different choices of the background template, see Table \ref{fluxes}. Those fluxes have the order the same as in Fig.\ \ref{spectra}.
} \label{fig:model_test}
\end{figure}

In addition to the described above tests, we have performed an additional test related to the possible uncertainties in the knowledge of exact shape and normalization of the spectra of point-like sources in the region. Namely, instead of modelling the spectra of the sources as best-fit power laws with free normalisation, we consider the model in which the spectral shapes in each energy bin are given by the best-fit models (which are for some sources more complex than power laws) from the 3FGL catalogue \citep{3rdcat}. All parameters except the normalisations were assumed to be fixed to their catalogue values. Then, in order to avoid possible contributions from sources beyond the ROI, we additionally included into the model the sources from the region larger by $10\degr$ than the ROI with all the parameters fixed to their catalogue model. The obtained results are consistent with the ones presented in the main part of the paper and are shown Fig.~\ref{fig:model_test}. The obtained fluxes are somowhat lower, and the significances of the flux difference between the soft and hard state are now lower. We note, however, that the 3FGL models were obtained from the fits at $E>100$~MeV and may not correctly reproduce the data at lower energies. Therefore, we are not sure which version is closer to the actual spectra. Overall, our results here confirm the robustness of the performed analysis and the detection of Cyg X-1 in both spectral states at energies $\lesssim$80~MeV.

\label{lastpage}


\begin{thebibliography}{}

\bibitem[\protect\citeauthoryear{Abdo et al.}{2009}]{fermi} 
Abdo, A. A., et al., 2009, Sci, 326, 1512

\bibitem[\protect\citeauthoryear{Acero et al.}{2016}]{acero16} 
Acero F., et al., 2016, ApJS, 223, 26 

\bibitem[\protect\citeauthoryear{Ackermann et al.}{2012}]{fermi_back} Ackermann M., et al., 2012, ApJ, 750, 3 

\bibitem[\protect\citeauthoryear{Aharonian \& Vardanian}{1985}]{av85} 
Aharonian F.~A., Vardanian V.~V., 1985, Ap\&SS, 115, 31 

\bibitem[\protect\citeauthoryear{Albert et al.}{2007}]{magic} 
Albert J., et al., 2007, ApJ, 665, L51 

\bibitem[\protect\citeauthoryear{Barthelmy et al.}{2005}]{barthelmy05}
Barthelmy S. D. et al., 2005, Space Sci. Rev., 120, 143

\bibitem[\protect\citeauthoryear{Blandford \& K{\"o}nigl}{1979}]{bk79} 
Blandford R.~D., K{\"o}nigl A., 1979, ApJ, 232, 34 

\bibitem[\protect\citeauthoryear{Bradt, Rothschild \& Swank}{Bradt et al.}{1993}]{brs93}
Bradt H. V., Rothschild R. E., Swank J. H., 1993, A\&AS, 97, 355

\bibitem[\protect\citeauthoryear{Brocksopp et al.}{1999}]{brocksopp99}
Brocksopp C., Fender R. P., Larionov V., Lyuty V. M., Tarasov A. E., 
Pooley G. G., Paciesas W. S., Roche P., 1999, MNRAS, 309, 1063

\bibitem[\protect\citeauthoryear{Caballero-Nieves et al.}{2009}]{cn09} 
Caballero-Nieves S.~M., et al., 2009, ApJ, 701, 1895 

\bibitem[\protect\citeauthoryear{Coppi}{1999}]{coppi99} 
Coppi P.~S., 1999, in Poutanen J., Svensson R., eds., ASP Conf.\ Ser.\ Vol.\ 161, High Energy Processes in Accreting Black Holes. Astron.\ Soc.\ Pac., San Francisco, p.\ 375

\bibitem[\protect\citeauthoryear{Di Salvo et al.}{2001}]{disalvo01} 
Di Salvo T., Done C., \.Zycki P. T., Burderi L., Robba N. R., 2001, ApJ, 547, 1024

\bibitem[\protect\citeauthoryear{Done, Gierli{\'n}ski \& Kubota}{Done et al.}{2007}]{dgk07} 
Done C., Gierli{\'n}ski M., Kubota A., 2007, A\&ARv, 15, 1 

\bibitem[\protect\citeauthoryear{Dubus}{2013}]{dubus13} 
Dubus G., 2013, A\&ARv, 21, 64 

\bibitem[\protect\citeauthoryear{Dubus, Cerutti, \& Henri}{2008}]{dch08} 
Dubus G., Cerutti B., Henri G., 2008, A\&A, 477, 691 

\bibitem[\protect\citeauthoryear{Dubus, Cerutti \& Henri}{Dubus et al.}{2010}]{dch10} 
Dubus G., Cerutti B., Henri G., 2010, MNRAS, 404, L55

\bibitem[\protect\citeauthoryear{Fermi-LAT Collaboration}{2015}]{3rdcat} 
Fermi-LAT Collaboration, 2015, ApJS, 218, 23

\bibitem[\protect\citeauthoryear{Frontera et al.}{2001}]{frontera01} 
Frontera F., et al., 2001, ApJ, 546, 1027 

\bibitem[\protect\citeauthoryear{Gierli{\'n}ski et al.}{1999}]{gierlinski99} 
Gierli{\'n}ski M., Zdziarski A.~A., Poutanen J., Coppi P.~S., Ebisawa K., Johnson W.~N., 1999, MNRAS, 309, 496 

\bibitem[\protect\citeauthoryear{Jackson}{1972}]{jackson72} 
Jackson J.~C., 1972, Nat.\ Phys.\ Sci., 236, 39 

\bibitem[\protect\citeauthoryear{Jourdain, Roques \& Malzac}{Jourdain et al.}{2012}]{jrm12} 
Jourdain E., Roques J.-P., Malzac J., 2012, ApJ, 744, 64

\bibitem[\protect\citeauthoryear{Laurent et al.}{2016}]{laurent16}
Laurent P., Gouiffes C., Rodriguez J., Chambouleyron. V., 2016, Gamma-Ray Astrophysics in Multi-Wavelength Perspective, 11th INTEGRAL Conference, \url{www.sron.nl/integral2016-programme}

\bibitem[\protect\citeauthoryear{Levine et al.}{1996}]{levine96}
Levine A. M., Bradt H., Cui W., Jernigan J. G., Morgan E. H.,

\bibitem[\protect\citeauthoryear{Malyshev, Zdziarski \& Chernyakova}{Malyshev et al.}{2013}]{mzc13} 
Malyshev D., Zdziarski A.~A., Chernyakova M., 2013, MNRAS, 434, 2380 (MZC13)

\bibitem[\protect\citeauthoryear{Markwardt et al.}{2005}]{m05} 
Markwardt C.~B., Tueller J., Skinner G.~K., Gehrels N., Barthelmy S.~D., Mushotzky R.~F., 2005, ApJ, 633, L77 

\bibitem[\protect\citeauthoryear{Matsuoka et al.}{2009}]{matsuoka09} 
Matsuoka, M. et al., 2009, PASJ, 61, 999

\bibitem[\protect\citeauthoryear{Mattox et al.}{1996}]{mattox96} 
Mattox J.~R., et al., 1996, ApJ, 461, 396 

\bibitem[\protect\citeauthoryear{McConnell et al.}{2002}]{mcconnell02} 
McConnell M.~L., et al., 2002, ApJ, 572, 984 (M02)

\bibitem[\protect\citeauthoryear{Orosz et al.}{2011}]{orosz11} 
Orosz J.~A., McClintock J.~E., Aufdenberg J.~P., Remillard R.~A., Reid 
M.~J., Narayan R., Gou L., 2011, ApJ, 742, 84

\bibitem[\protect\citeauthoryear{Pooley \& Fender}{1997}]{pf97} 
Pooley G.~G., Fender R.~P., 1997, MNRAS, 292, 925

\bibitem[\protect\citeauthoryear{Poutanen \& Coppi}{1998}]{pc98} 
Poutanen J., Coppi P.~S., 1998, PhST, 77, 57 

\bibitem[\protect\citeauthoryear{Poutanen \& Vurm}{2009}]{pv09} 
Poutanen J., Vurm I., 2009, ApJ, 690, L97 

\bibitem[\protect\citeauthoryear{Reid et al.}{2011}]{reid11} 
Reid M.~J., McClintock J.~E., Narayan R., Gou L., Remillard R.~A., Orosz 
J.~A., 2011, ApJ, 742, 83 

\bibitem[\protect\citeauthoryear{Rodriguez et al.}{2015}]{rodriguez15} 
Rodriguez J., et al., 2015, ApJ, 807, 17 

\bibitem[\protect\citeauthoryear{Rushton}{2009}]{rushton09} 
Rushton A. P., 2009, PhD thesis, University of Manchester

\bibitem[\protect\citeauthoryear{Rushton et al.}{2011}]{rushton11} 
Rushton A., et al., 2011, Proceedings of Science, 10th EVN Symposium, 061

\bibitem[\protect\citeauthoryear{Stawarz et al.}{2004}]{stawarz04} 
Stawarz {\L}., Sikora M., Ostrowski M., Begelman M.~C., 2004, ApJ, 608, 95 

\bibitem[\protect\citeauthoryear{Stirling et al.}{2001}]{stirling01} 
Stirling A.~M., Spencer R.~E., de la Force C.~J., Garrett M.~A., Fender R.~P., Ogley R.~N., 2001, MNRAS, 327, 1273 

\bibitem[\protect\citeauthoryear{Vladimirov et al.}{2011}]{GALPROP} Vladimirov A.~E., et al., 2011, CoPhC, 182, 1156, arXiv:1008.3642

\bibitem[\protect\citeauthoryear{Walter \& Xu}{2017}]{walter16}
Walter R., Xu M., 2017, A\&A, in press, arXiv:1706.07962

\bibitem[{Wilks(1938)}]{wilks38}
Wilks S.~S., 1938, The Annals of Mathematical Statistics, 9, 60

\bibitem[\protect\citeauthoryear{Zanin et al.}{2016}]{zanin16} 
Zanin R., Fern{\'a}ndez-Barral A., de O{\~n}a-Wilhelmi E., Aharonian F., Blanch O., Bosch-Ramon V., Galindo D., 2016, A\&A,  596, A55 (Z16)

\bibitem[\protect\citeauthoryear{Zdziarski, Pooley \& Skinner}{Zdziarski et al.}{2011a}]{zps11} 
Zdziarski A.~A., Pooley G.~G., Skinner G.~K., 2011a, MNRAS, 412, 1985

\bibitem[\protect\citeauthoryear{Zdziarski et al.}{2011b}]{z11b} 
Zdziarski A.~A., Skinner G.~K., Pooley G.~G., Lubi{\'n}ski P., 2011b, MNRAS, 416, 1324

\bibitem[\protect\citeauthoryear{Zdziarski, Lubi{\'n}ski \& Sikora}{Zdziarski et al.}{2012}]{zls12} 
Zdziarski A.~A., Lubi{\'n}ski P., Sikora M., 2012, MNRAS, 423, 663

\bibitem[\protect\citeauthoryear{Zdziarski et al.}{2014a}]{z14a} Zdziarski A.~A., Stawarz {\L}., Pjanka P., Sikora M., 2014a, MNRAS, 440, 2238 

\bibitem[\protect\citeauthoryear{Zdziarski et al.}{2014b}]{z14b} 
Zdziarski A.~A., Pjanka P., Sikora M., Stawarz {\L}., 2014b, MNRAS, 442, 3243 (Z14)

\bibitem[\protect\citeauthoryear{Zdziarski et al.}{2016}]{zpr16} 
Zdziarski A.~A., Paul D., Osborne R., Rao A. R., 2016, MNRAS, 463, 1153

\bibitem[\protect\citeauthoryear{Zi{\'o}{\l}kowski}{2005}]{ziolkowski05}
Zi{\'o}{\l}kowski J., 2005, MNRAS, 358, 851

\bibitem[\protect\citeauthoryear{Zi{\'o}{\l}kowski}{2014}]{ziolkowski14}
Zi{\'o}{\l}kowski J., 2014, MNRAS, 440, L61 

\end{thebibliography}
\end{document}